# Heterogeneous MPSoCs for Mixed Criticality Systems: Challenges and Opportunities

Mohamed Hassan *Member, IEEE*
mohamed.hassan@ieee.org
Intel Corp.

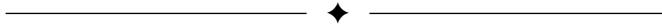

**Abstract**—Due to their cost, performance, area, and energy efficiency, MPSoCs offer appealing architecture for emerging mixed criticality systems (MCS) such as driverless cars, smart power grids, and healthcare devices. Furthermore, heterogeneity of MPSoCs presents exceptional opportunities to satisfy the conflicting requirements of MCS. Seizing these opportunities is unattainable without addressing the associated challenges. We focus on four aspects of MCS that we believe are of most importance upon adopting MPSoCs: theoretical model, interference, data sharing, and security. We outline existing solutions, highlight the necessary considerations for MPSoCs including both opportunities they create and research directions yet to be explored.

## 1 INTRODUCTION

Real-time systems are those systems, whose proper behaviour depends not only on their functionality but also on their response time. Until recently, real-time systems have been limited to safety-critical domains such as avionics and spacecrafts. However, with the emanating Cyber-Physical Systems (CPS) and Internet of Things (IoT) revolution, real-time systems are becoming ubiquitous in many emerging domains. Examples include transportation such as smart vehicles, infrastructures such as power grids, healthcare such as implantable devices, and industrial environment such as robots. These domains pose two new major aspects that did not traditionally exist in real-time systems: the mixed criticality nature of its software applications, and the Multiple-Processor System-on-Chip (MPSoC) architecture of its hardware components.

1) *Mixed Criticality Systems (MCSs).* These domains are no longer solely hosting isolated safety-critical tasks. Instead, they execute various tasks with different criticalites, where the criticality of a task is determined based on the consequences of the failure to meet its requirements. For instance, Figure 1 illustrates a subset of the tasks embedded on a modern vehicle. Tasks such as the Anti-lock Braking System (ABS), the steering, and the engine control units are of high-criticality. Meeting the timing requirements of those tasks (historically known as Hard-Real Time (HRT) tasks) is a life-safety condition. Other tasks such as the infotainment system and the connectivity box (such as internet, radio, WiFi,..etc.) are of low criticality in the sense that they do not require strict timing guarantees. Instead, their proper functionality requires a high average-case performance. A third class of tasks contains tasks with medium criticality, known as Soft-Real Time (SRT) tasks, such as the navigation system and the instrument cluster in a vehicle. They require a predictable execution time, which is not as strict as higher-critical tasks, as well as a reasonable average-case performance. The number of criticality levels is domain specific and is not limited to three. For instance, the DO-178C avionics standard defines five levels of assurance, while the ISO-26262 defines four automotive safety integrity levels.

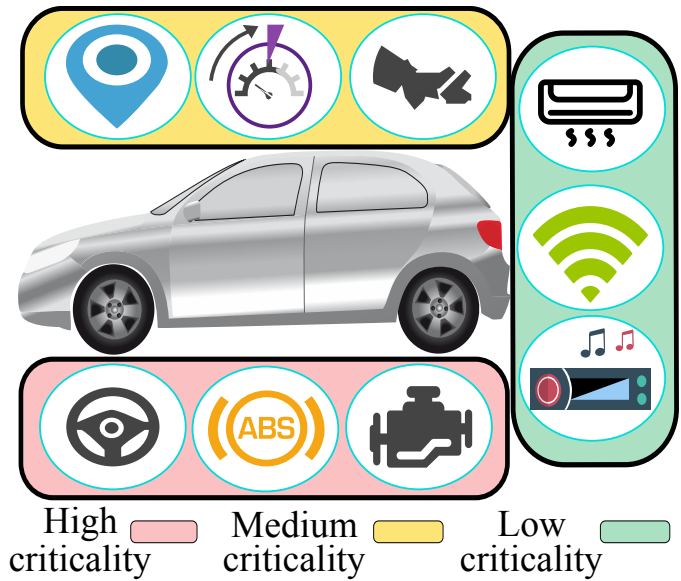

Fig. 1: Examples of tasks running on a modern vehicle.

2) **MPSoCs.** MPSoCs are appealing platforms for emerging MCS domains primarily due to the benefits they provide in cost, area, power-consumption, and performance compared to traditional computing systems. In addition, heterogeneous MPSoCs allow for customized solutions to increase these benefits. The main intuition is that designing a single Processing Element (PE) to meet conflicting requirements of MCS tasks is inefficient due to the limited cost, area, and battery budgets of MCS. Contrarily, designing custom PEs towards meeting those requirements has already proved its efficaciousness in current SoCs. Examples include specialized Digital Signal Processors (DSPs), cipher, and multimedia PEs. In fact, one of the early motives of evolving MPSoCs was the real-time, low area, and low-power demands from embedded systems [1]. The envision for MCS is that a task with a particular criticality can be scheduled in an expedient core with the appropriate level of hardware predictability. Figure 2 de-lineates an example of such heterogeneous MPSoC architecture. In the near future, MPSoCs are expected to be used in all embedded systems domains [2], [3]. To make this a reality, researchers made sincere efforts to provide MPSoCs tailored for safety-critical tasks (e.g. [2]–[4]). Companies started to develop MPSoCs that include dictated real-time processing units such as the Zynq UltraScale+ MPSoC [5] from Xilinx, and the heterogeneous Nona-Core SoC from Renesas [6]. Safety standards are also slowly shifting towards considering multiple PEs. For instance, the AUTOSAR standard from the automotive industry released a guide to deploy software tasks onto multi-core architectures in a recent revision [7].

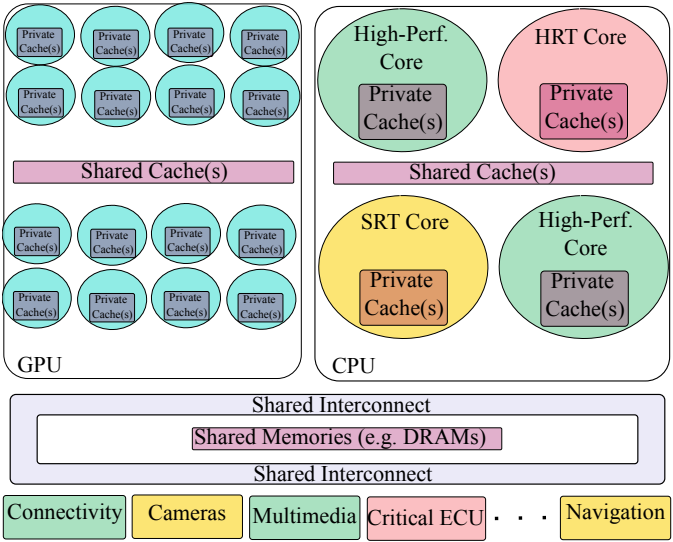

Fig. 2: Examples of tasks running on a modern vehicle.

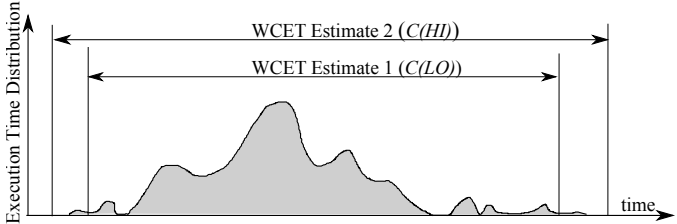

Fig. 3: Different WCET estimates.

These two aspects together (MCS and MPSoCs) of emerging embedded systems bring out a number of challenges that has to be carefully repelled. The focus of this paper is to highlight those challenges, the proposed solutions in literature to address them, and the open issues yet to be addressed. We limit our discussion to four aspects of MCS: theoretical modelling, timing interference, data sharing, and security.

## 2 MCS MODEL

**Current Model.** As identified by Vestal in his constitutional paper [8], the MCS model differs from the traditional real-time task model because of the uncertainty in considered WCET. Basically, the computed WCET of a task is an estimate calculated using extensive experimental testing and/or static analysis methods. Hence, based on the accuracy and pessimism levels of these methods, different estimates may exist (Figure 3). The higher the criticality of a task is, the more pessimistic its WCET estimates are. This observation resulted in representing the WCET as a function in the criticality level, $C(l)$. The majority of MCS papers consider a model of only two CLs, $LO$ and $HI$ [9]. Each task has $C(LO)$ and $C(HI)$, where $C(LO) \leq C(HI)$. The system operates initially in a normal mode, where it considers the $C(LO)$ of each task and both higher- and lower-critical tasks utilize the hardware resources. Run-time techniques are used to monitor execution times of running tasks. If a critical task exceeds its $C(LO)$, the system switches to a degraded mode, where it suspends all lower-critical tasks and considers the $C(HI)$ of the higher-critical ones. This dynamic migration between various modes is a key characteristic of MCS as compared to single-criticality systems.

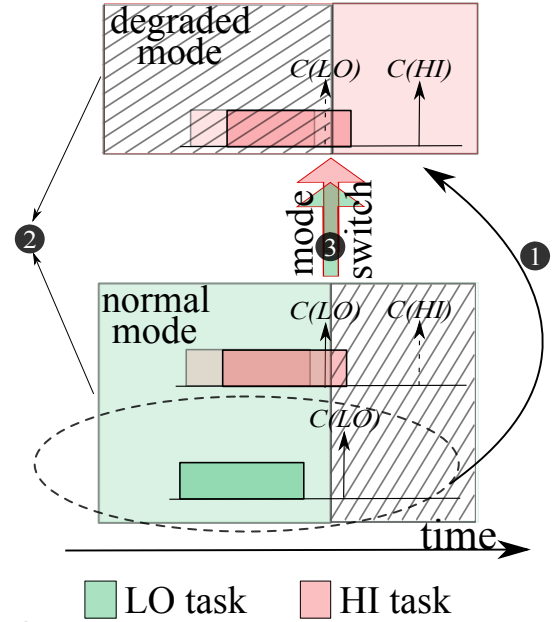

Fig. 4: Current MCS Model and its issues.

### 2.1 Issues with the Current Model.

There exist number of issues with approaches adopting this simplified widely-considered model. Figure 4 highlights the three issues we believe are of most importance in MPSoCs.

1) **Suspension.** Upon switching to the degraded mode, no guarantees are given to lower-critical tasks. The dual-criticality model deems lower-critical as non-critical; hence, there are no consequences of suspending them. Nevertheless, in systems with multiple CLs such as in the ISO-26262 standard, which has four ASILs A (lowest) to D (highest), suspension of tasks of ASIL A may be acceptable, while suspension of tasks of ASIL C may be prohibitively unacceptable solution as it may result in safety issues [10]. Different approaches than suspension has to be considered. The approach in [11] changes the model such that $C(HI) \leq C(LO)$ for lower-critical tasks. Accordingly, upon switching to the degraded mode, instead of abandoning them, their guarantees are only degraded. Another approach is followed by [12], where instead of directly switching the mode and suspending lower-critical tasks, the memory service guarantees of those tasks are degraded to reduce the interference on the higher-critical to accommodate for the increase in the execution time. These approaches consider a system-wide mode switch, where all the system components and tasks migrate to the new mode.

**MPSoC Reconsiderations.** In an MPSoC, there may be no need to deploy such full-system mode migration. Assume a scenario where a task, $\tau_i$ running on the medium-criticality (SRT) core in Figure 2 such that it exceeds its depicted $C$ for the current mode, say because of a soft fault or a temperature increase in the SRT core. There exist opportunities to keep other non-interfering cores running the same set of tasks (i.e. no effective mode switching), while switching only the necessary core(s). Other technique can be migrating tasks to "more-predictable" cores to avoid more switching at all. For the exemplified scenario, tasks running on that particular SRT-core can be migrated to an HRT-core (if possible) upon the monitored increase in $\tau_i$'s $C$. So, a set of run-time decisions now exist thanks to the heterogeneity nature of MPSoCs. Such MPSoC-related opportunities are yet to be explored.

2) **Number of CLs and Sources of Uncertainty.** Restricting the model to only two criticalities is not sufficient to meet industry standards, which define up to five levels as aforestated. It may seem that extending approaches that consider this model to more than two criticalities is straightforward. However, in most cases it is not. For instance, the suspension issue discussed in the first point is an outcome of a dual-criticality model, in which lower-critical tasks are deemed non-critical. For systems with multiple levels, different approaches than suspension have to be considered.

**MPSoC Reconsiderations.** The heterogeneous nature of MPSoCs has a direct effect on the number of CLs. The standard model assumes that the uncertainty in WCET does not come from the system itself; rather, it comes from our inability to measure (or compute) it with complete confidence [9]. Although the later part of this assumption still holds for MPSoCs, the former does not. Yes, the WCET estimate in MPSoCs is still a function of our confidence-level of the used tools. However, we argue that the architecture of MPSoCs originates uncertainties as well. In traditional single-core or Symmetric Multiple-Processor (SMP) architectures, which core (or cores) is executing a task does not affect its measured execution time. However, in a heterogeneous MPSoC (such as in Figure 2), the decision of which cores are used to execute a task directly affects the level of certainty in its WCET. For instance, an HRT core is usually simple in terms of micro-architecture with almost no implemented architectural optimizations. This is necessary to allow for high-level of analysability, which leads to tight WCETs for safety-critical tasks. Contrarily, a high-performance core usually deploys speculative optimizations such as out-of-order execution and branch prediction. As a result, the confidence level in a task's WCET when it runs on an HRT core definitely differs from the case when the same task runs on a high-performance core. The interdependency dilemma that requires investigation here is that the WCET estimates become a function in the task-to-core mapping, which is part of the scheduling algorithm that relies on these estimates as inputs.

3) **Overheads.** Monitoring tasks and switching between running modes engross high overheads. However, to simplify the scheduling problem of MCS, most approaches ignore these overheads. Although this may be a theoretically acceptable assumption since these overheads are implementation-related, a practical adoption of these approaches in industry mandates careful quantification of these overheads. Recently, a few efforts have been proposed to bridge this gap [13], [14]. While the former [13] focuses on single-core, the latter [14] evaluates multi-core platforms. Both efforts consider the implementation of a subset of proposed scheduling mechanisms; thus, more studies are required to identify implementation-related issues of theory-based novel MCS scheduling techniques.

**MPSoC Reconsiderations.** Deployment of MCS onto MPSoCs requires addressing the scalability challenges associated with these scheduling and monitoring techniques. Further, mode switching in MPSoCs may incur task migrations or reassignment of heterogeneous cores to tasks; thus, the effects of these decisions on the switching overhead need to be quantified. On the other hand, MPSoCs open the door for customized solutions. For instance, dedicating a PE for the run-time monitoring possibly helps in faster detection of exceptional events, therefore enabling the system to react in a timely manner. The architecture of this PE can be further tailored to optimize the behaviour of the monitoring techniques. Specialized PEs are widely used by current MPSoCs, which usually dictate PEs for security, connectivity, data signal processing, and other tasks.

## 3 SHARED RESOURCES: TIMING INTERFERENCE

In MPSoCs, different PEs in the system interfere with each other, while competing to access memory resources that are shared amongst them. As Figure 2 depicts, these shared resources include interconnects, on-chip caches, and off-chip Dynamic Random Access Memories (DRAMs). This interference is a challenge for real-time systems because operations of one core affect the temporal behaviour of other cores, which complicates the timing analysis of the system.

Since the aforementioned MCS model originally evolved for single-core systems, most of proposed approaches adopting it do not incorporate these interferences in their scheduling or analysis [15]. Experiments show that memory interferences can contribute up to $300\%$ to the WCET [16], while the memory bus interference can solely increase the WCET up to $44\%$ [17]. Consequently, it is of unavoidable necessity to account for these interferences for MPSoC MCS. There exist proposals to address this interference in multi-core MCS at the interconnect (e.g. [12], [15]), the shared cache (e.g. [18], [19]), and the shared DRAM (e.g. [20]–[22]). However, most of these approaches consider SMP architectures and do not account for the heterogeneity of MPSoCs.

**MPSoC Reconsiderations.** Bounding the timing interference in MPSoCs is a burdensome goal for many reasons.

1) The interference exaggerates with the increase in number of PEs competing on the shared resources.
2) Each type of PEs has its own memory access behaviour, which complicates the analysis, thus leading to more pessimism. For instance, data-intensive PEs such as multimedia and DSP processors can saturate system queues by their memory requests if not carefully arbitrated. Thus, unlike most of the current approaches, a requirement- and a criticality-aware arbitration is a must to deliver differential service to PEs.
3) Understanding the architectural details of shared resources (such as the interconnect and the memory hierarchy) is inevitable to derive realistic bounds.

On the other hand, MPSoCs provide unique opportunities that do not exist in SMPs. Efforts to investigate these opportunities will facilitate the deployment of MCS onto MPSoCs. Examples of research directions include:

1) Which memory levels should be shared amongst which cores to meet various requirements of MCS. For instance, Does the GPU share the last-level cache with the GPU?
2) How to distribute the cache architecture? Would implementing a Non-Uniform Cache Architecture (NUCA), which is a norm in many-core systems, be an adequate approach for MCS (e.g. helping in achieving different levels of isolation)?
3) Usually, MPSoCs integrate different types of on-chip memories such as hardware-managed caches and software-managed Scratch-Pad Memories (SPMs). Most of the currently available approaches focus on a single type. Similarly, MPSoCs support different types of available off-chip memories such as Double Data Rate (DDR), Graphics DDR (GDDR), Low-Power DDR (LPDDR). Investigating the co-operation of these types is also worth investigating.

## 4 SHARED RESOURCES: SHARED DATA

One of the most challenging burdens for computer architects is to maintain correctness of shared data stored in memory hierarchies of multiple-PEs platforms, which is known as *cache coherence*. Although cache coherence has been extensively investigated for conventional performance-oriented platforms, embedded systems introduce new challenges from the predictability perspective. For instance, empirical studies show that the data interference and the coherence effects can make the parallel execution of an application $3.87\times$ slower than its sequential execution [23], while the worst-case coherence latency exhibits quadratic growth with increasing number of PEs [24]. Real-time community has introduced various solutions to address this problem, which we categorize into three approaches identifying the applicability of each approach to MPSoCs.

1) **Prevention**. This approach avoids the problems resulting from data sharing by completely disallowing it through enforcing complete isolation between tasks. At the shared cache, mechanisms such as strict cache partitioning and colouring are used [25]. At the DRAM level, bank privatization is utilized to uniquely map tasks to different banks [20]. Isolation is an attractive solution as it simplifies the analysis and minimizes timing interference, while ensuring data correctness. However, data isolation suffers from three limitations. 1) It adopts the independent-task model, thus disabling any communication amongst tasks. 2) It may result in a poor memory or cache utilization. For instance, a task can keep evicting its cache lines if it reaches the maximum of its partition size, while other partitions may remain underutilized. 3) It does not scale with increasing number of cores. For example, the number of cores in the system has to be less than or equal to the number of DRAM banks to be able to achieve isolation at DRAM.

**MPSoC Reconsiderations.** It might be acceptable for traditional real-time task models to assume isolation in order to achieve uniprocessor-equivalence [26] in multiprocessor platforms, thus allowing the reuse of maturely developed scheduling approaches for uniprocessors. However, with the emerging technologies that continuously adopt new functionalities, complete isolation seems to be a prohibitively costly solution. Considering the automotive applications in Figure 1, they utilize data collected by various sensors to conduct the appropriate act. This data is usually shared and used by applications with different-criticalities. For instance, the brake sensors are utilized by both the high-critical ABS, and the driver assistance and cruise control tasks which are of medium-criticality [27]. In addition, with the massive concurrency of MPSoCs, solutions preventing simultaneous running of dependent (i.e. data sharing) tasks are becoming evidently ill-suited as they diminish the performance gains of MPSoCs due to the aforementioned three limitations. Accordingly, researchers recognized that in order to have any practical impact, scheduling techniques must permit data sharing [24], [28] and the following two approaches are recently proposed to solve data sharing problems without enforcing isolation.

2) **Shared-data aware scheduler**. This approach combines operating system techniques, profiling, and hardware performance counters to envision the effects of data sharing. Accordingly, the scheduler is constructed such that it minimizes those effects [23], [28]. For example, if two tasks share data, a simple solution is to schedule them such that they do not simultaneously run on different PEs. This can be achieved either by postponing the execution of one of them, or mapping them to the same PE [28].

**MPSoC Reconsiderations.** These scheduling-based solutions are promising since they address the shared data problem, while not enforcing isolation. However, some of these solutions have limited applicability to MPSoCs. For example, given the high level of parallelism in MPSoCs, mapping dependent tasks to same core may not be a viable solution as it limits performance gains. Similarly, collecting run-time readings from performance counters may not be costly effective in terms of overheads given the large number and heterogeneity of PEs in MPSoCs.

3) **Predictable hardware cache coherence**. A recent work [24] manages shared data in real-time systems through deploying a hardware cache coherence protocol. Identifying the sources of unpredictability due to coherence interference, this work promotes certain invariants to maintain towards allowing simultaneous and predictable accesses to shared data. These invariants are satisfied by augmenting the classic Modify-Share-Invalidate (MSI) protocol with transient coherence states, and minimal architectural changes. The advantages of this approach is that programmers do not need to explicitly manage coherence of shared data in the application. In addition, it does not require any modifications to existing scheduling algorithms.

**MPSoC Reconsiderations.** This approach does not address coherence in MCS. In addition, this approach, along with all the discussed approaches, consider SMPs. Effects of heterogeneity on these approaches is not investigated yet. For example, how coherence operates across different types of PEs? One coherence protocol might not fit all types and allowing for different coherence protocols can be a better solution for MCS if the interaction between these protocols is carefully designed.

## 5 SECURITY OF MPSOC MCS

Security is one of the biggest challenges encountering researchers and engineers of CPS. The more ubiquitous the CPS become, the more concerning their security is. To exemplify, various vulnerabilities have been reported in industrial SCADA systems [29] and smart vehicles [30], [31].

**MPSoC Reconsiderations.** Three aspects make the development of secure MPSoC MCS a burden challenge.
1) **The Cyber-physical nature of MCS.** MCS in many emerging domains interact with the physical world, which makes them CPS. Embedded components in these CPS manage sensitive tasks; therefore, any security breach could lead to catastrophic consequences. These consequences range from revealing personal information (e.g. from wearable devices) to a global threat (e.g. compromising a nuclear plant). Consequently, ensuring the security of these systems is a first-class mission. In addition, the interaction with the physical world allows for threats that did not exist in traditional computing systems. For example, researchers successfully managed to gain access to locked cars by only eavesdropping a single signal from the original remote keyless entry unit of the car [30].

TABLE 1: Opprotunities and Challenges.

| Aspect | Challenges | Opportunities/Research directions |
|---|---|---|
| Theoretical Model | 1. WCET becomes a function of task-to-core mapping<br>2. Scalability of monitoring techniques and switching overheads<br>3. Effects of heterogeneity on task migration upon switching | 1. Spatial/partial instead of system-wide mode switching<br>2. Alternatives to mode switch such as migrating tasks to more-predictable cores<br>3. Customizable solutions: e.g. dedicating specialised PE for execution-time monitoring |
| Timing Interference | 1. Large number of PEs<br>2. PEs have different memory behaviors<br>3. Hard to derive tight bounds | 1. Which levels should be shared amongst which PEs for MCS?<br>2. Distribution of cache architecture. e.g, uniform or NUCA?<br>3. Different types of on- (caches and SPMs) and off-chip memories (DDR, GDDR, LPDDR) |
| Data Sharing | 1. massive parallelism deems isolation ill-suited<br>2. Different memory patterns complicates the analysis<br>3. Scheduling-based approaches are hard to adopt | 1. Different PEs have different transaction sizes and BW requirements<br>2. Different memory types with different features (high bandwidth vs. low latency)<br>3. Possibility for different approaches for each PE type |
| Security | 1. CPS entails possible catastrophic threats<br>2. New heterogeneity-exploiting threats<br>3. New vulnerabilities because of Shared components between different criticalities | 1. Identifying MPSoCs specific vulnerabilities<br>2. Developing cost- and performance-effective methodologies to face threats<br>3. Adopting security as a first-class citizen in designing MPSoCs for MCS |

2) **The heterogeneity of MPSoCs.** On one hand, each PE of MPSoC has different characteristics and can even be an intellectual proprietary of a third-party entity. Accordingly, the security problems of each of these PEs are inherited. On the other hand, these PEs share system components and interact with each other. This opens the door for new across-PEs threats. Accordingly, threats and vulnerabilities in MPSoCs are harder to analyse, detect, and assess compared to traditional systems. To exemplify, the well-known Stuxnet attack exploited the authentication of the Siemens programmable logic controller to access a Windows machine [29].

3) **The shared hardware components in MPSoCs.** Historically, security was not considered as a concern for MCS because isolation was a design aspect of these systems, where each task (or group of tasks with same criticality) is running on a PE (or a partition of PEs) that is completely isolated from other PEs (or partitions). As a consequence, sensitive tasks that require high levels of security is isolated from non-secure tasks. However, in MPSoCs, this is not the case. Different PEs, and hence tasks, share hardware components and isolation is considered a costly solution as previously explained. Again, this creates new potential threats. To exemplify, researchers were able to control sensitive (considered secure) ECUs such as the engine control in a Jeep Cherokee car by compromising the (considered insecure) radio unit because the radio unit shares the CAN network with these ECUs [31]. To address these challenges, three research directions are necessary towards secure deployment of MPSoC MCS: 1) identifying new vulnerabilities of MPSoCs that did not exist in traditional platforms, 2) developing cost- and performance- effective methodologies to prevent or mitigate them, and 3) adopting security as a first-class citizen in designing MPSoCs for MCS (secure-by design concept).

## 6 CONCLUSION

We argue that MPSoCs will be soon the dominating platform for emerging embedded systems domains. Despite the tremendous benefits and opportunities they provide, certain challenges have to be addressed and new research directions need to be explored. Four aspects are of great importance upon deploying MCS on MPSoCs: theoretical modelling, timing interference, data sharing, and security. For each of these aspects, Table 1 highlights both the remarkable challenges and the opportunities that MPSoCs uniquely create. These challenges and opportunities can be back traced to three characteristics of MPSoCs: 1) large number of PEs, 2) heterogeneity of PEs, 3) different types of shared resources amongst PEs. We believe that seizing these opportunities and addressing associated challenges will enable enormous advances for MCS applications.